\documentclass[journal,twoside]{IEEEtran}
\usepackage{gensymb}
\usepackage{textcomp}
\usepackage[mathlines,switch,pagewise]{lineno}
\usepackage{cite}
\ifCLASSINFOpdf
  \usepackage[pdftex]{graphicx}
\else
  \usepackage[dvips]{graphicx}
\fi
\usepackage[cmex10]{amsmath}
\usepackage{xfrac}
\usepackage{amsfonts}
\usepackage[linesnumbered, lined, commentsnumbered, ruled]{algorithm2e}
\usepackage{array}
\usepackage{multicol}
\usepackage{colortbl}
\usepackage{footnote}
\ifCLASSOPTIONcompsoc
  \usepackage[caption=false,font=normalsize,labelfont=sf,textfont=sf]{subfig}
\else
  \usepackage[caption=false,font=footnotesize]{subfig}
\fi
 \usepackage{dblfloatfix}
\usepackage{url}
\usepackage[colorlinks=true, citecolor=green, linkcolor=red, urlcolor=blue]{hyperref}
\usepackage[hyperpageref]{backref}
\usepackage{stackengine}
\usepackage{color}
\usepackage{xcolor}

\newcolumntype{N}{>{\centering\arraybackslash}m{2.9cm}}
\newcolumntype{M}{>{\centering\arraybackslash}m{2.4cm}}
\newcommand{\white}[1]{{\textcolor{white}{#1}}}
\linenumbers

\begin{document}
\nolinenumbers
\bstctlcite{IEEEexample:BSTcontrol}
\title{Robust Segmentation of Cell Nuclei in 3-D Microscopy Images}

\author{Sundaresh~Ram,~\IEEEmembership{Member,~IEEE,}
        and~Jeffrey~J.~Rodr\'iguez,~\IEEEmembership{Senior Member,~IEEE}% <-this % stops a space
\thanks{Manuscript received Month xx 20xx; revised Month xx 20xx; accepted Month xx 20xx. \textit{(corresponding author: Sundaresh Ram)}}% <-this % stops a space
\thanks{Sundaresh Ram is with the Department of Radiology, and the Department of Biomedical Engineering, University of Michigan, Ann Arbor, MI, 48109 USA (e-mail: sundarer@umich.edu).}
\thanks{Jeffrey J. Rodr\'iguez is with the Department
of Electrical and Computer Engineering, The University of Arizona, Tucson,
AZ, 85721 USA (e-mail: jjrodrig@email.arizona.edu).}}


% \markboth{}%
% {}
\markboth{IEEE Transactions on Biomedical Circuits and Systems,~Vol.~xx, No.~x, Month~20xx}%
{Ram and Rodriguez: Robust Segmentation of Cell Nuclei in 3-D Microscopy Images}
% {Ram \MakeLowercase{\textit{et al.}}: Robust Segmentation of Cell Nuclei in 3-D Microscopy Images}
% after the title page when using the two side option.

% column for its text to clear the IEEE pub id mark.


\maketitle

\begin{abstract}
Accurate segmentation of 3-D cell nuclei in microscopy images is essential for studying the nuclear organization, gene expression, and cell morphodynamics. Current image segmentation methods are challenged by the complexity and variability of microscopy images and often over-segment or under-segment the cell nuclei. Thus, there is a need to improve segmentation accuracy and reliability, as well as the level of automation. This paper presents an automated algorithm for segmentation of 3-D cell nuclei using the concepts of keypoint features, random walk, graph theory, and energy minimization as the foundation. We first use a seed detection algorithm to find a seed voxel for each individual cell nucleus. Next, using the concept of random walk on a graph we find the probability of all the voxels in the 3-D image to reach each of the seed voxels. We then generate a 3-D response image by combining these probabilities for each voxel and apply the marker controlled watershed transform on this response image to obtain an initial segmentation of the cell nuclei. Finally, we apply local region-based active contours to obtain final segmentation of the cell nuclei. The advantage of using such an approach is its ability to accurately segment highly textured cells having inhomogeneous intensities, and varying shapes and sizes. The proposed algorithm was compared with three other nucleus segmentation algorithms for segmentation accuracy using overlap measure, Tanimoto index, Rand index, F-score, and Hausdorff measure. Quantitative and qualitative results show that our algorithm provides improved segmentation accuracy compared to existing algorithms.
\end{abstract}

% Note that keywords are not normally used for peer review papers.

% \fi
%
% For peer review papers, this IEEEtran command inserts a page break and
\IEEEpeerreviewmaketitle

\section{Introduction}
\label{sec:intro}
\IEEEPARstart{T}{he} 3-D segmentation of cells in fluorescent in situ hybridization (FISH) images is an important first task for studying the underlying nuclear organization \cite{ram12}. The use of FISH probes that fluoresce at particular wavelengths makes possible the visualization of the entire nucleus as well as sub-nuclear structures \cite{shiels07, ram10}. Traditional genetic studies have sought to determine the actual function of individual genes within cells and how their linear sequences encode specific physical traits, while more recent genomic and system-level methods now allow biologists to query how all genes function together at any given time in a single cell, cell populations, tissues and whole organisms. Genetic studies have now converged to show what profound effects the 3-D organization of the genome can have, depending on the genetic information being read from the DNA molecule \cite{misteli07}. The idea that the physical location of a gene within the nucleus is important for its normal function comes from several observations suggesting that the 3-D organization of chromatin within the eukaryotic nucleus is non-random \cite{takiazawa08}. This non-random organization allows for the placement of genes into specific compartments that facilitate either gene activation or gene repression. This suggests that the localization of gene sequences within the nucleus impacts their expression, and mis-localization of genomic regions could lead to mis-regulation of gene expression. Thus, there is significant interest in accurately segmenting 3-D nuclei so as to identify the location of sub-nuclear structures within them.

Cell segmentation is a widely studied topic with numerous available algorithms \cite{chawla04, cheng09, dufour05, dzyubachyk10, kofahi10, lankton08, cli08, lankton082, li08, roysam06, ram12-2, zanella10, lee13, dewan14, ram17, ram20, ram21}, but new methods continue to be investigated due to the many challenges posed by new applications. One of the main difficulties of 3-D cell segmentation in microscopy images is due to the way they are imaged. Since the 3-D data is captured by imaging one focal plane at a time, it consists of a stack of 2-D slices (planar images). Focusing on one plane at a time leads to partially imaged nuclei in slices, sectioning of nuclei at odd angles, and since each 2-D slice has a fixed thickness it results in partially superposed cell nuclei in the image slices. As a result of these imaging limitations, we observe a set of image objects that significantly differ from the ideal round objects that we would expect. There are various other sources of difficulties that arise in 3-D cell segmentation: closely clustered cells which are touching and overlapping in 3-D \cite{ram13}, cells of varying shapes and sizes, accidental and non-specific staining \cite{ram21,ram22}, the highly textured nuclei due to variable chromatin structure, images where the contrast between the foreground cells and the background is low \cite{lee13}, fluctuating intensities, the presence of spectral unmixing errors in processed multispectral images \cite{dewan14}, and the imaging noise \cite{malladi20}, especially for fluorescence data. Also, the fact that these cells are large and span multiple images calls for an algorithm which is not only effective, but at the same time computationally reasonable.

Dufour \emph{et al.} \cite{dufour05} proposed a cell segmentation and tracking scheme on the human parasite Entamoeba histolytica and MDCK cells. Their method uses multiple active surfaces with or without edges, coupled by a penalty for overlaps and a volume conservation constraint that improves outlining of cell boundaries. This method works well when the gradient information varies smoothly from the center of the cell towards its edges but fails to accurately segment highly textured cells. Also, initialization of the active surface is very important as a wrong initialization can lead to errors in segmentation.

Li \emph{et al.} \cite{li07, li08} developed a segmentation algorithm based on gradient vector flow tracking on C. elegans embryo and HeLa (human cancer) cell nuclei images. Their method is composed of three key steps: (1) generate a diffused gradient vector flow field by an elastic deformation transformation; (2) perform a gradient flow tracking procedure to attract points to the basin of a ``sink" (the center of the nucleus); and (3) separate the image into small regions, each containing one nucleus and nearby peripheral background, and perform local adaptive thresholding in each small region to extract the cell nucleus from the background. This method requires that the centers of the nuclei are brighter than the nearby surrounding regions and thus will not perform well on textured cell images, where the center may not be the only bright region within a nucleus.

Dzyubachyk \emph{et al.} \cite{dzyubachyk10} developed a scheme to segment cell nuclei from HeLa and Chinese hamster ovary (CHO) cell images. Their method consists of several improvements and extensions to the coupled active surfaces algorithm developed by Dufour \emph{et al.} \cite{dufour05}. There are two key improvements: (1) application of the radon transform to “decouple” the active surfaces of touching cells by means of a separating plane, in order to apply a stopping criterion to each level set function separately; and (2) incorporation of a modified energy scheme based on a non-linear partial differential equation. If there are touching cells in the initial frame, then this method does not separate them, leading to errors in the subsequent frames. Also, clustered cells are always partitioned via a plane, which may not be the actual boundary separating them.

Cheng \emph{et al.} \cite{cheng09} proposed a method to segment and separate clustered nuclei in neuronal cell images. Their method makes use of Dufour's coupled active surfaces algorithm \cite{dufour05} to find the image foreground regions, an adaptive H-minima transform to find shape-based markers, and a marking function based on the outer distance transform to separate the clustered nuclei. This method works well when the gradient information varies smoothly from the center of the cell towards its edges and depends on the initialization of active surfaces, as wrong initialization can lead to errors in segmentation. Additionally, for highly textured clustered cells the delineation process leads to errors in finding the exact borders of the clustered cells.

Al-Kofahi \emph{et al.} \cite{kofahi10} developed a method to segment cell nuclei from histopathology images that gained a lot of popularity. Their algorithm consists of four steps: (1) extraction of the image foreground using graph-cuts binarization; (2) detection of nuclear seed points using multiscale Laplacian-of-Gaussian filtering constrained by distance-map adaptive scale selection; (3) initial segmentation using the seed points; and (4) refinement using a second graph-cuts algorithm incorporating the method of alpha expansions and graph coloring to reduce computational complexity. For this method to work well, the input images are required to have a bimodal histogram, and for large polyploid cells this method finds multiple seeds within one nucleus.

In this paper, we present a new automated algorithm for segmentation of 3-D FISH images obtained via confocal microscopy, applied to the study of ovarian germline nurse cells of Drosophila melanogaster. First, we employ a seed detection/marker extraction algorithm to find exactly one seed/marker per nucleus proposed previously in \cite{ram16}. Next, we find the probabilities of each voxel in the 3-D image to reach each seed/marker within the image by treating the image as a graph and using the concept of random walk on graph. Once we get the various probabilities associated with each voxel we formulate a 3-D response image by combining these probabilities and run the marker controlled watershed algorithm on this response image to obtain an initial segmentation of the 3-D cell nuclei. Finally, we make use of 3-D active contours as a post-processing step to optimize the initial segmentation results based on shape and smoothness and also correct for other errors. Segmentation overlap measure \cite{sonka07}, Tanimoto index \cite{duda01}, \cite{jimenez06}, F-score \cite{ram18}, Rand index \cite{unnikrishnan07} and Hausdorff distance \cite{huttenlocher93} are used to measure the performance of nucleus segmentation.

The main contributions of the proposed segmentation approach are three-fold. First, by applying a seed detection algorithm that uses the principle of radial symmetry nature of cells, we are able to accurately identify one seed per nucleus with varying nuclei sizes within one image. This is very important as earlier segmentation methods tend to under-segment a small nucleus that is in the vicinity of a large nucleus, thereby causing errors for further analysis. Second, we generate a 3-D response image where the image intensity within the cells decreases monotonically from the inner core of the cell towards the cell borders, and we run the watershed transform on this image thereby ensuring that there is no over-segmentation of the highly textured cells, a case generally observed in the prior segmentation approaches. Third, by applying a local active contours technique, we are able to prune away the unwanted background voxels included as foreground within the cells and smooth the contours in 3-D along the cell nuclei borders.
%--------------------------------------------- FIGURE - 1 ---------------------------------------------------%
\begin{figure*}[!t]
\subfloat[]{\includegraphics[width= 1.75in, height=1.75in]{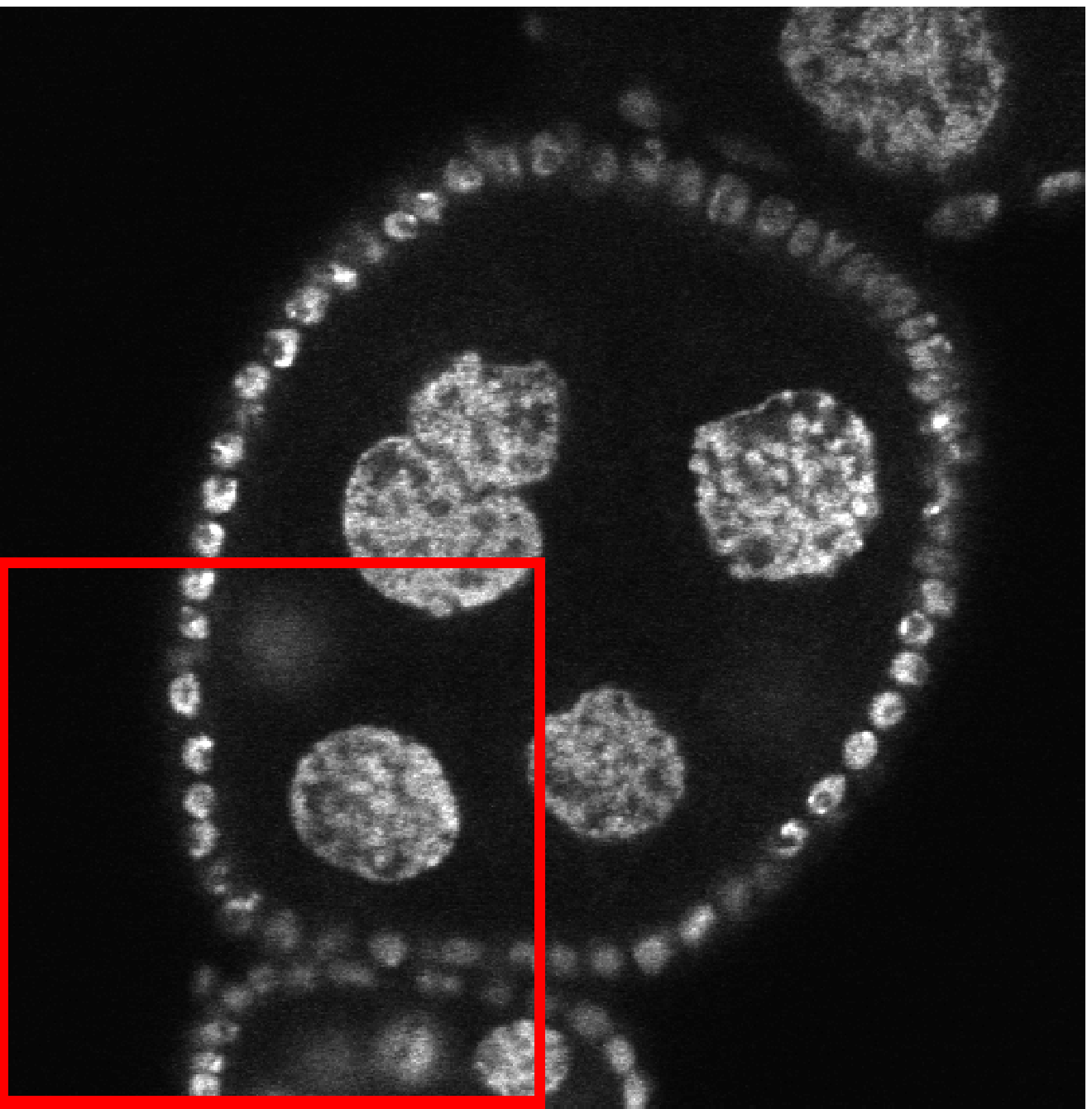}}\hfill
\subfloat[]{\includegraphics[width= 1.75in, height=1.75in]{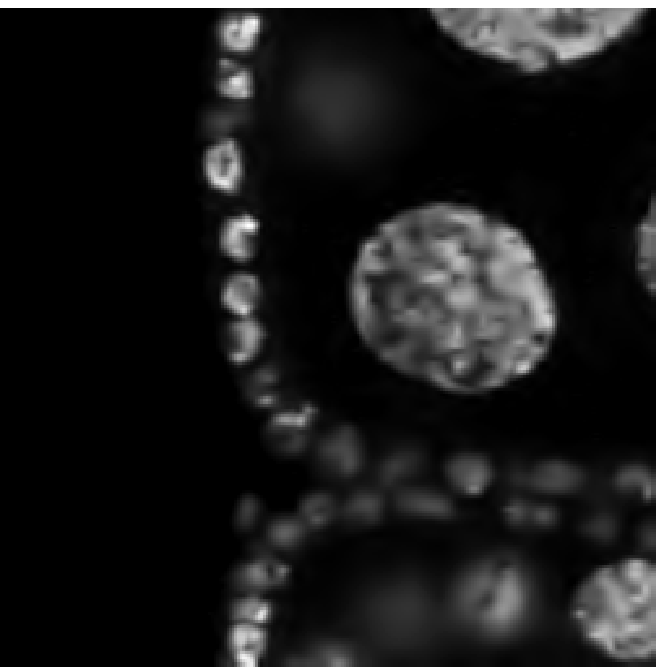}} \hfill
\subfloat[]{\includegraphics[width= 1.75in, height=1.75in]{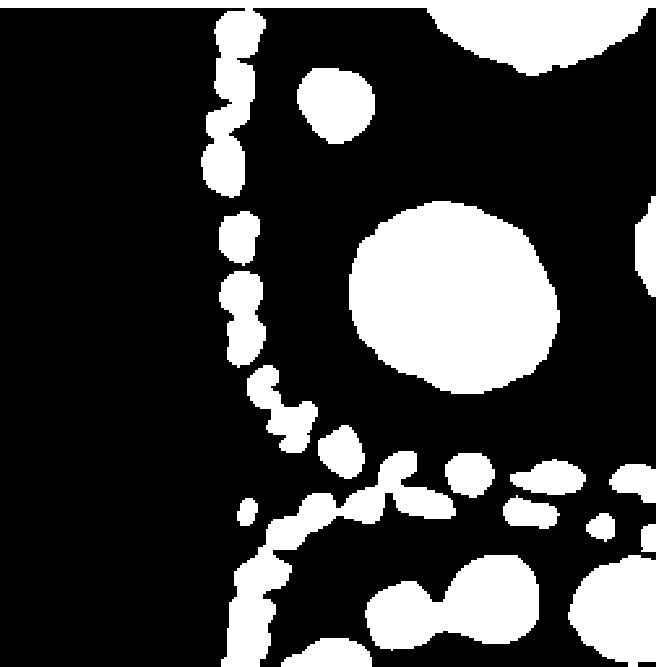}}\hfill
\subfloat[]{\includegraphics[width= 1.75in, height=1.75in]{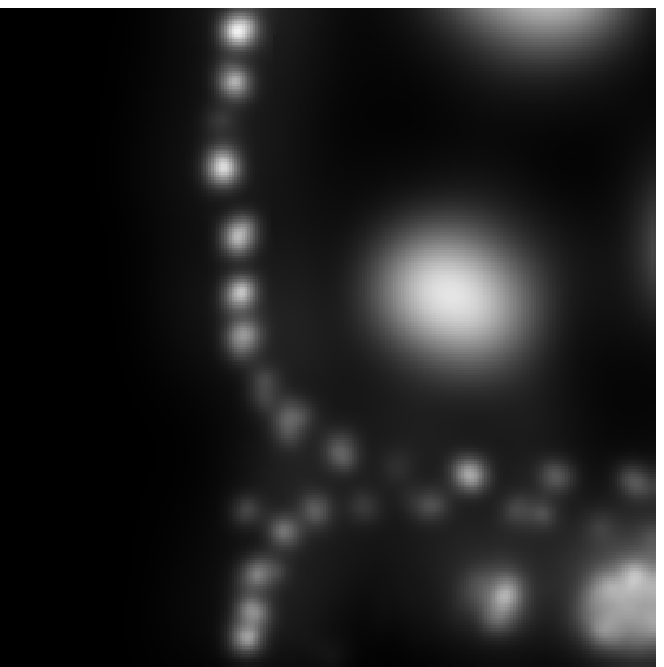}}\vspace{-1.5mm}\\
\subfloat[]{\includegraphics[width= 1.74in, height=1.75in]{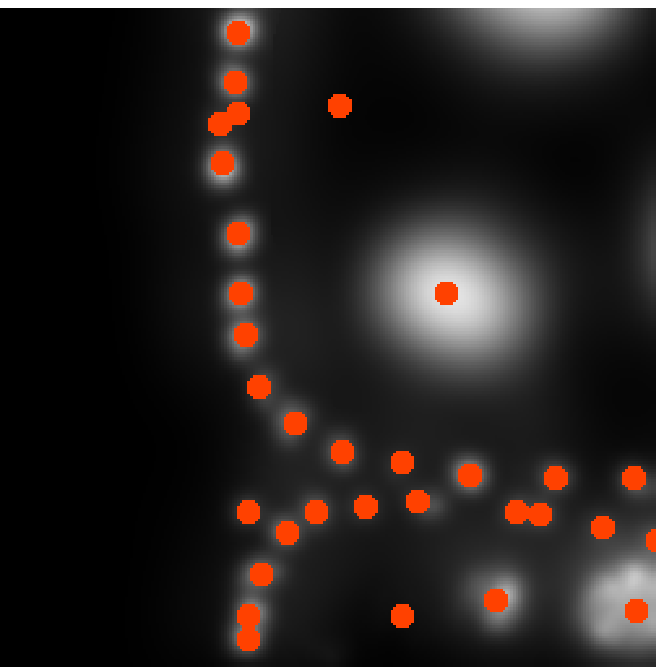}}\hfill
\subfloat[]{\includegraphics[width= 1.73in, height=1.75in]{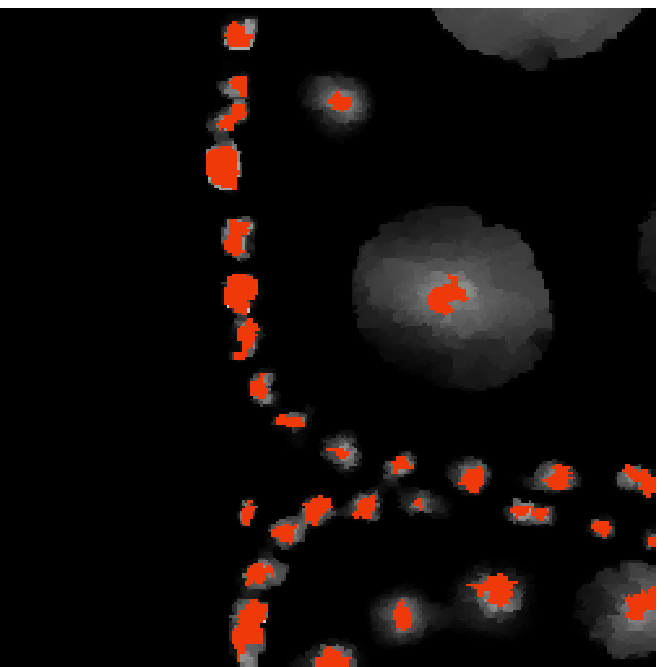}} \hfill
\subfloat[]{\includegraphics[width= 1.73in, height=1.75in]{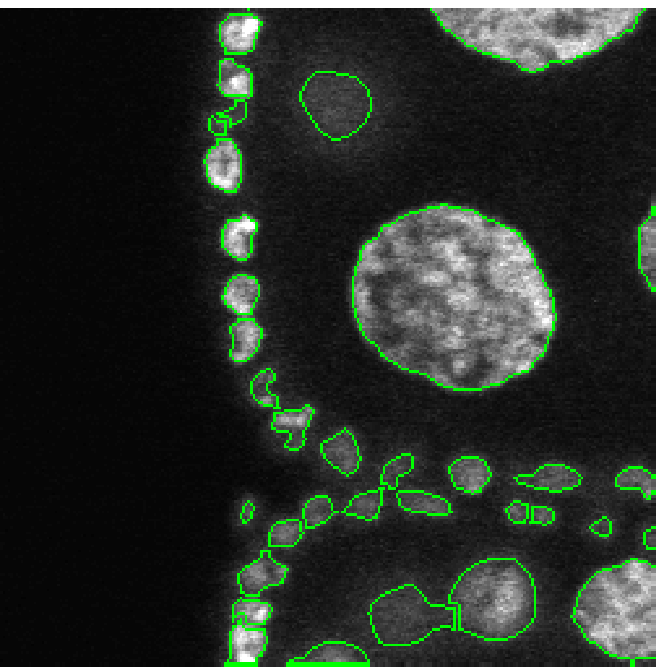}}\hfill
\subfloat[]{\includegraphics[width= 1.73in, height=1.75in]{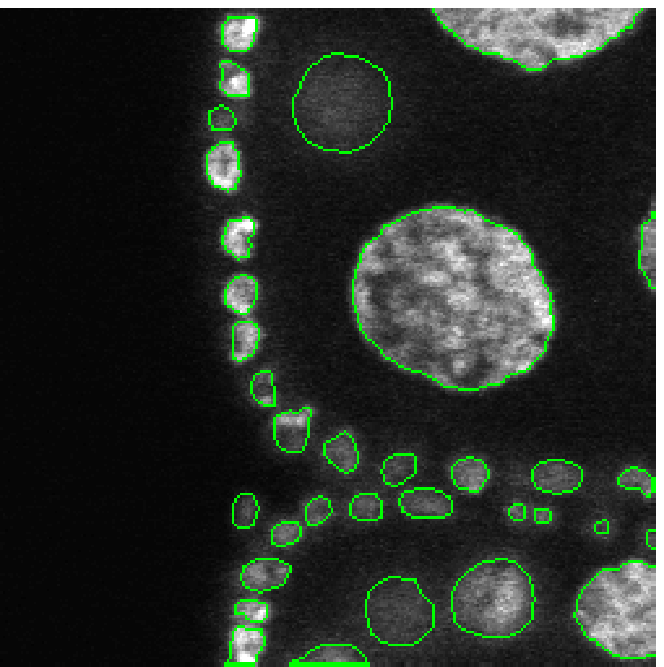}}
\caption{The steps of  nucleus segmentation on a single 2-D slice. The steps are shown for the region marked with a red box. (a) Original 2-D slice. (b) Denoised. (c) After thresholding. (d) After applying FRST. (e) Detected seeds superimposed on the FRST. (f) After growing the initial seed regions. (g) Initial segmentation of the cell nuclei. (h) Final segmentation of cell nuclei after post-processing.}
\label{fig2}
\vspace{-3mm}
\end{figure*}
%----------------------------------------------------------------------------------------------------------------

\section{Material}
\label{sec:mat}

\subsection{Nuclear Staining and Hybridization}
\label{subsec:fishpro}

Conventional FISH protocols are cumbersome and time consuming. Below, we briefly describe the FISH protocol procedure as in \cite{ram12}, which was primarily adapted from \cite{derburg00}.  About 1-2 day old female flies were flattened on yeast with male flies at $25\,^{\circ}\mathrm{C}$ for 2 days.  Five ovary pairs were dissected in Grace’s medium (Invitrogen) and fixed within 15 minutes. The ovaries were stained with $10\,\mathrm{ng/mL}$ $4^{'},6$-diamidino-2-phenylindole (DAPI) in 2X SSCT for 10 minutes and then washed twice with 2X SSCT for 10 minutes each. Ovarioles were mounted in VECTASHIELD.  Pieces of number $1\frac{1}{2}$ coverslips were used as spacers between the actual coverslip and slide to prevent flattening of the egg chambers.

\subsection{Image Acquisition}
\label{subsec:imgaqui}

We used a Zeiss LSM 510 Meta confocal microscope (Carl Zeiss, Inc.) to acquire the images of the DAPI stained cells. The nucleus regions were obtained using an LP 420 DAPI filter at a wavelength of 405 $\mathrm{nm}$. Images were acquired with a Plan-Neofluar lens with magnification of 40X, numerical aperture = 1.3, and a voxel size of 0.31 $\mathrm{\mu m}$ in the x and y-direction and 1 $\mathrm{\mu m}$ in the z-direction with automatic focusing. A photomultiplier tube absorbed the signals emitted at specific wavelengths and transformed them into an intensity image $I \in \mathbb{R}^{3}$ according to the absorbed light. We adjusted the gain and offset of the photomultiplier tubes in order to prevent the signals from getting over-saturated or under-saturated.

%-------------------------------------------------------------------------------------------------------------------%
\section{Methods}
\label{sec:mthd}

The ovarian germline of the Drosophila melanogaster consists of two types of cells, namely ``nurse cells" and ``follicle cells" \cite{spradling93}. The follicle cells are smaller than the nurse cells and surround the nurse cells in an ellipsoidal fashion in 3-D. Both the follicle cells and nurse cells are brighter than the background region. Fig. \ref{fig2}(a) shows a single slice of a 3-D data set where the nurse cells are surrounded by smaller follicle cells lying along an elliptical ring.
%We define the background between the nurse cells and follicle cells as the inner background and the background between the follicle cells and the borders of the image as the outer background.
 
\subsection{Seed Detection}
\label{subsec:det}

Detection of seeds or markers is a widely researched topic, and many methods exist. Some of the popular methods to extract markers include the distance transformation, h-dome transformation, extended minima transformation, Laplacian of Gaussian (LoG) filtering, and morphological reconstruction. Recently, Ram and Rodriguez \cite{ram13, ram16} proposed a method to detect individual cell nuclei in microscopy images. This method uses the multiscale variance stabilizing transform \cite{zhang08} to suppress the noise from the images, histogram thresholding to locate all the foreground nuclei, and the fast radial symmetric transform and non-maximum suppression to find the center of each individual nucleus. We use this method to detect the initial seeds/markers of our nuclei in 3-D as it is size-invariant and is able to detect cells of varying sizes within an image. Figs. \ref{fig2}(b)-(e) show the results after each step of the seed detection algorithm on a representative 2-D slice from one of our 3-D data sets.

\subsection{Initial Segmentation}
\label{subsec:iniseg}

Using the markers obtained from the seed detection step, we apply the 3-D random walker algorithm developed by Grady \cite{grady06}. The random walker algorithm is a graph-based algorithm. A graph consists of a pair $G = (V, E)$ with vertices (nodes) $v \in V$ and edges $e \in E \subseteq V \times V$. An edge $e$ spanning two vertices, $v_{i}$ and $v_{j}$, is denoted by $e_{ij}$. A weighted graph assigns a weight $w : E \rightarrow \mathbb{R}$ to each edge. The weight of an edge $e_{ij}$ is denoted by $w(e_{ij})$ or simply $w_{ij}$. The degree of a vertex is $d_{i} = \sum_{j} w(e_{ij})$ summed over all edges $e_{ij}$ incident on $v_{i}$. In order to interpret $w_{ij}$ as the bias affecting a random walker's choice of path, we require that $w_{ij} > 0$. We also assume that our graph is simple, undirected and connected. To represent the image structure in terms of edge weights, we define a Gaussian weighting function given by
\begin{eqnarray}
w_{ij} = \exp \left\{ -\beta(q_{i} - q_{j})^{2} \right\}
\label{eqn:key1}
\end{eqnarray}
where $q_{i}$ is the image intensity at voxel $i$. A random walk on a graph is a simple Markov process that begins at some vertex $v_{i}$ and moves to a neighboring vertex $v_{j}$ with a probability $p_{ij}$ proportional to the weight of the corresponding edge, i.e., $p_{ij} = w_{ij}/d_{i}$. Given an image with the seed voxels $v_{j} \in V_{P}$ obtained from the seed detection step, one can analytically determine the probability $x_{i}$ that a random walker starting at a given unlabeled voxel $v_{i} \in V_{U}$ will first reach a prelabeled voxel. 

Computing these random walker probabilities directly is computationally impractical. However, the random walker probabilities can be found by solving the equivalent combinatorial Dirichlet problem, which involves finding a harmonic function subject to its boundary values \cite{grady06, ram13-2}. A harmonic function is a function that satisfies the Laplace equation. We define a combinatorial Laplacian matrix $L$ whose elements $L_{ij}$ for vertices $v_{i}$ and $v_{j}$ are defined as
\begin{eqnarray*}
L_{ij}&=& \left\{ \begin{array}{l l}
d_{i}& \quad \text{if}~i = j,\\
-w_{ij}& \quad \text{if}~v_{i}~\text{and}~v_{j}~\text{are adjacent nodes,}\\
0& \quad \text{otherwise}
\end{array} \right.
\end{eqnarray*}
We also define an $m\times n$ edge-node incidence matrix $A$, where $m$ and $n$ are number of edges and vertices, respectively, whose elements $A_{e_{ij}v_{k}}$ for each edge $e_{ij}$ and vertex $v_{k}$ are defined as
\begin{eqnarray*}
A_{e_{ij}v_{k}} &=& \left\{\begin{array}{l l}
+1& \quad \text{if}~i = k,\\
-1& \quad \text{if}~j = k,\\
~0& \quad \text{otherwise}\\
\end{array} \right.
\end{eqnarray*}
The matrix $A$ acts as a combinatorial gradient operator and the matrix $A^{\top}$ as a combinatorial divergence. The isotropic combinatorial Laplacian is given by $L = A^{\top}A$. We define the $m\times m$ constitutive matrix $C$ as the diagonal matrix with the weights of each edge along the diagonal. The combinatorial formulation of the Dirichlet problem is then given by \cite{grady06}
\begin{eqnarray}
D[x] = \frac{1}{2}(Ax)^{\top}C(Ax) = \frac{1}{2}x^{\top}Lx = \frac{1}{2}\sum\limits_{e_{ij} \in E} w_{ij}\left(x_{i} - x_{j}\right)^{2} \hspace{-20pt} \nonumber\\
\label{eqn:dirich}
\end{eqnarray}
where $x$ is a vector of length $\left| V \right|$, representing a probability value at each vertex of the graph $G$. Since $L$ is positive semidefinite, the only critical points of $D[x]$ will be minima. We partition the vertices into two sets: $V_{P}$ representing the prelabeled nodes or detected seeds and $V_{U}$ representing the unlabeled nodes. Then $V_{P}\cup V_{U} = V$ and $V_{P}\cap V_{U} = \emptyset$. We can decompose (\ref{eqn:dirich}) as
\begin{eqnarray}
\hspace{-2mm} D[x_{_U}] \hspace{-1mm} & = & \hspace{-1mm}\frac{1}{2}\left[{\begin{array}{cc}
x_{_{P}}^{T} & x_{_{U}}^{T} \\
\end{array}}\right]\left[{\begin{array}{cc}
L_{P} & B\\
B^{T} & L_{U}\\
\end{array}}\right]\left[{\begin{array}{c}
x_{_{P}}\\
x_{_{U}}\\
\end{array}}\right] \nonumber \\
\hspace{-3mm} & = & \hspace{-1mm} \frac{1}{2}\left(x_{_{P}}^{T}L_{P}x_{_{P}} + 2x_{_{U}}^{T}B^{T}x_{_{P}} + x_{_{U}}^{T}L_{U}x_{_{U}}\right)
\label{decomp}
\end{eqnarray}
where column vectors $x_{_P}$ and $x_{_U}$ correspond to the random walker probabilities at the prelabeled and unlabeled voxels, respectively. For each unlabeled voxel $v_{u} \in V_{U}$, we denote by $x_{u}^{j}$ the probability that a random walker, starting at $v_{u}$ will first reach the $j^{th}$ prelabeled voxel, $v_{j} \in V_{P}$.

Differentiating $D[x_{_U}]$ with respect to $x_{_U}$ and finding critical points yields
\begin{eqnarray}
L_{U}x_{_{U}} = -B^{\top}x_{_{P}}
\label{derivative}
\end{eqnarray}
Equation (\ref{derivative}) represents a system of linear equations with $|V_{U}|$ unknowns. The solution to the combinatorial Dirichlet problem may then be found by solving
\begin{eqnarray}
L_{U}x^{j} = -B^{\top}p^{j}
\end{eqnarray} 
for each label $j$ or
\begin{eqnarray}
L_{U}X = -B^{\top}P
\end{eqnarray}
for all the labels, where $X$ has $K$ columns taken by each $x^{j}$ and $P$ has $K$ columns taken by each $p^{j}$, i.e.
\begin{eqnarray*}
X &=& \left[x^{j},~ 0 < j \leq K\right]\\
P &=& \left[p^{j},~ 0 < j \leq K\right].
\end{eqnarray*}
Here $p^{j}$ is a $|V_{M}| \times 1$ vector for each label $j$ given by $p^{j} = \left[\delta_{p}^{j},~ 0 < j\leq K\right]^{\top}\hspace{-1mm}, \delta_{i}^{j}$ is the Kronecker delta function. Since the probabilities at any node will sum to unity, i.e.
\begin{eqnarray}
\sum\limits^{K}_{j = 1} x_{u}^{j} = 1 \quad \forall~~~ v_{u} \in V_{U}
\end{eqnarray} 
only $K-1$ sparse linear systems need to be solved, where $K$ is the total number of detected seeds.

We now consider a zero 3-D image $R \in \mathbb{R}^{3}$, called the response image of the same size as the original 3-D image $I$ and mark the prelabeled voxels detected using the seed detection step in Section \ref{subsec:det} with labels $s$, where $s \in \mathbb{Z}, 0 < s \leq K$. Once we obtain the $K$-tuple vector of probabilities for each unlabeled voxel, we combine this vector of probabilities into one value in the response image $R$:
\begin{eqnarray}
R(v_{u}) = \prod\limits^{K}_{j = 1} x_{u}^{j} & \quad \text{if}~v_{u} \in V_{U}
\label{res:img}
\end{eqnarray}
This response image $R$ will have maximum values in the areas where the probabilities are equal for every $0 < j \leq K$, i.e. when an unlabeled voxel has equal probability to reach any of the prelabeled voxels. Since the probability of an unlabeled voxel to reach any prelabeled voxel decreases as we move away from this prelabeled voxel, there is a ridge formation within each nucleus region in the response image $R$ generated by (\ref{res:img}). We grow the markers/seeds of each nucleus obtained using the FRST transform in section \ref{subsec:det} in 3-D, until we reach the ridge of that particular nucleus in the response image $R$, thereby obtaining a bigger seed region for each nucleus in 3-D. Fig. \ref{fig2}(f) shows the grown seed regions of each nucleus across the single 2-D slice. We set these grown 3-D seed regions and the background to negative infinity value on the response image $R$, negate the resulting image and apply the 3-D watershed algorithm to this negated image to obtain the accurately segmented nuclei in 3-D with the nuclei delineated. Fig. \ref{fig2}(g) shows the contours of a single slice overlaid onto the nucleus regions within the image.

\subsection{Post Processing}
\label{subsec:postproc}

In section \ref{subsec:iniseg} the result of the segmentation using the 3-D watershed transformation leads to detection of jagged boundaries around the nucleus. It is necessary discard the background voxels wrongly labeled as foreground, if any, and smoothen the jagged boundaries around each nucleus in order to obtain a complete segmentation. In the recent years, active contours have shown to provide a very appropriate framework for this purpose, so we employ a 3-D active contour method for solving this problem. One major drawback of using active contours has been with regard to poor initialization, leading to erroneous solutions. We do not face this problem, as our initialization for each nucleus is the boundary found by the 3-D watershed transformation, which is very close to the true boundary of the nucleus to be determined. Since the goal is to find a true boundary for each textured 3-D nucleus and there are densely populated areas, it is more effective to use a region-based active contour scheme which looks at the local statistics of the region rather than the global statistics. There are active contour methods in the literature \cite{lankton08, cli08} that look at local region statistics for image segmentation. Lankton \emph{et al.} \cite{lankton08} developed a localized region-based active contour for image segmentation, which considers local statistics along the evolving contour and is capable of segmenting objects with heterogeneous feature profiles.  Li \emph{et al.} \cite{cli08} proposed a region-based active contour model for image segmentation that draws upon intensity information in local regions at a controllable scale in order to segment textured regions within an image. This model uses a data fitting term that includes a kernel function using which information about local regions can be extracted to guide the motion of the evolving contour. One can choose any of the above methods for obtaining the final segmentation results. In this work, we use Lankton's method with minor modifications in our algorithm to obtain accurate segmentation of the 3-D nuclei. This method considers local regions around each point in the initial contours and finds the local energies corresponding to these regions so as to aid evolution of the contour. To compute these local energies, each point on the evolving contour is considered separately, and the local neighborhoods of this point are divided into local interiors and local exteriors. The local energies are then optimized by moving each point in the evolving curve in a direction so as to minimize the energy computed in its local interiors and local exteriors \cite{lankton08}.

We denote $I$ as the input image with domain $\Omega$, and let $S = \{x \in \Omega \,|\, \phi (x) = 0\}$ be the closed contour represented as the zero level set of a signed distance function $\phi$. The signed distance function $\phi$ is given by
\begin{eqnarray}
\phi (x) &=& \left\{\begin{array}{l l}
\inf\limits_{y \in \Omega^{c}} d(x,y)& \text{if} ~ x \in \Omega \\
-\inf\limits_{y \in \Omega} d(x,y)& \text{if} ~ x \in \Omega^{c} \\
\end{array} \right.
\end{eqnarray}
where $d(x,y)$ denotes the Euclidean distance function. We specify the interior of $S$ as a set of points $x$ using the approximation of the smoothed Heaviside function as
\begin{eqnarray}
H(\phi(x)) = \left\{\begin{array}{l l}
1,&  \phi (x) > \varepsilon \\
0,&  \phi (x) < -\varepsilon \\
\frac{1}{2} \left[ 1 + \frac{\phi}{\varepsilon} + \frac{1}{\pi} \sin \left(\frac{\pi \phi (x)}{\varepsilon}\right) \right],& \text{otherwise}\\ 
\end{array} \right. \hspace{-20pt} \nonumber \\
\end{eqnarray} 
Similarly, the exterior of $S$ is defined as $(1 - H(\phi(x)))$. The local area just around the evolving contour is specified by the derivative of $H(\phi(x))$, a smoothed version of the Dirac delta function given by
\begin{eqnarray}
\delta (\phi(x)) = \left\{\begin{array}{l l}
1,&  \phi (x) = 0 \\
0,&  \mid \! \phi (x) \!\mid < \varepsilon \\
\frac{1}{2 \varepsilon} \left[ 1 + \cos\left(\frac{\pi \phi (x)}{\varepsilon}\right) \right],& \text{otherwise}\\ 
\end{array} \right.
\end{eqnarray}
We introduce a second variable $y$. We will use $x$ and $y$ to denote two independent variables each representing a single point in $\Omega$. Using this notation, we introduce a function $B(x, y)$ in terms of a radius parameter $\Psi$ as
\begin{eqnarray}
B(x, y) &=& \left\{\begin{array}{l l}
1,&  \parallel \! x-y \!\parallel < \psi \\
0,&  \text{otherwise} \\
\end{array} \right.
\end{eqnarray}
where
 \begin{eqnarray}
\psi &=& \left\{\begin{array}{l l}
\Psi,&  \parallel \! x-y \!\parallel < \parallel \! x-\tilde{x} \!\parallel \\
\parallel \! x-\tilde{x} \!\parallel,&  \text{otherwise} \\
\end{array} \right.
\label{eqn:20}
\end{eqnarray}
and $\tilde{x}$ is a point on the nearby evolving contour. The function $B(x, y)$ is used to mask local regions corresponding to the point $x$. This function will take a value of 1 when the point $y$ is within a ball of radius $\psi$ centered at $x$ and $0$ otherwise. Using $B(x, y)$ we define an energy function $E$ in terms of a generic force function $F$ along with a regularization term to keep the curve smooth. The energy function $E$ is given by
\begin{eqnarray}
E(\phi) = \int_{\Omega_{x}} \delta \phi (x) \int_{\Omega_{y}} B(x, y)  F(I(y), \phi (y)) \,dy \,dx \nonumber \\ 
 + ~ \lambda \int_{\Omega_{x}} \delta \phi(x) \parallel \! \nabla \phi(x) \!\parallel \,dx  
\label{eqn:egy}
\end{eqnarray}
where $\lambda$ is a parameter that penalizes arc-length of the curve. By taking the first derivative of the energy in (\ref{eqn:egy}) with respect to $\phi$ we obtain the following equation:
\begin{eqnarray}
\frac{\partial \phi}{\partial t} (x) = \delta \phi(x) \int_{\Omega_{y}} B(x, y)  \partial_{\phi (y)} F(I(y), \phi (y)) \,dy \nonumber \\ 
~~~~ + ~ \lambda \delta \phi(x) ~ \text{div}\left( \frac{\nabla \phi (x)}{| \nabla \phi (x) |} \right)  
\end{eqnarray}
where $t$ is an artificial (evolution) time parameter.

We use the same local energies, the \emph{uniform modeling energy}, the \emph{means separation energy}, and the \emph{histogram separation energy} as computed in \cite{lankton08} to be inserted into the generic force function $F$ (see \cite{lankton08} for details of how each of these energies are computed and presented in the evolution equation). The convergence of the evolution function is computationally intensive and depends on the size of the local mask, in our case $B(x, y)$. Also, we need to re-initialize the signed distance function $\phi$ every few iterations until convergence. In order to accelerate the computation time, we use the sparse field method (SFM) proposed by Whitaker \cite{whitaker98}. SFM reduces the computational effort by considering just a single layer of active grid points for time integration --- i.e., it uses the lists of points that represent the zero level set as well as points adjacent to the zero level set. By using these lists and carefully moving points to and from the appropriate list, a very efficient representation of $\phi$ can be maintained. The detailed procedure about the implementation of SFM is explained in \cite{lankton082}. There are advantages of using SFM: 1) the speed of the curve updates is dependent only on the length of the curve and not on the entire image, and in our case it depends only on the local mask $B(x, y)$; 2) SFM does not require re-initialization, unlike the narrow band method \cite{li10}. Since we have good initialization and there are very few voxels that need to be thrown away as belonging to the background, the evolution equation converges very fast for each nucleus. Fig. \ref{fig2}(h) shows a single slice where the final contours are overlaid onto the nucleus regions after applying the 3-D active contours as a post-processing step. 

%-------------------------------------------------------------------------------------------------------------------%

\section{Results}
\label{sec:res}

We evaluate the proposed algorithm with respect to two different tasks: nucleus segmentation and nucleus detection. Overlap measure, Tanimoto coefficient, F-score, Rand index, and Hausdorff distance are used to measure the performance of nucleus segmentation. We compared our algorithm with three other algorithms developed by Cheng \emph{et al.} \cite{cheng09}, Dzyubachyk \emph{et al.} \cite{dzyubachyk10}, and Al-Kofahi \emph{et al.} \cite{kofahi10}. The optimal parameters for all the algorithms for our data were selected using the FROC analysis by varying one parameter at a time while keeping the others fixed and choosing that value of the parameter which maximized the area under the curve (AUC).

We used thirty eight 3-D data sets consisting of about 280 2-D slices per data set. The dimension of each 2-D slice is 504 $\times$ 512 pixels, and the gray scale dynamic range of each data set is 8 bits per pixel. A total of 8519 follicle cells and 584 nurse cells were labeled by an expert and considered for testing the segmentation accuracy.

\subsection{Algorithm Parameter Settings}
\label{subsec:algparam}

For all the datasets, the parameters of our proposed algorithm (Section \ref{sec:mthd}) were fixed to the following values after performing the FROC analysis. In the FRST transform used for seed detection, the radius parameters used were $n_{\min} = 2$ and 
\begin{eqnarray}
n_{\max} = 3 \times  D(x, y, z)
\end{eqnarray}
where $D(x, y, z)$ is the 3-D distance transform \cite{haralick92} of the binarized image, the variance of the 3-D Gaussian kernel ($G_{n}$) used was $\sigma^{2} = (n_{\min} + n_{\max})/2$, and the radial strictness parameter and the normalizing parameter were $\gamma = 2$ and $k_{n} = 10$, respectively. In the random walker algorithm the free parameter used in (\ref{eqn:key1}) is $\beta = 50$. In the post-processing step, the radius parameter in (\ref{eqn:20}) is $\Psi = 9$, and the parameter that penalizes the arc-length of the contour in (\ref{eqn:egy}) is $\lambda = 0.5$. For Cheng's method \cite{cheng09}, after the FROC analysis the active contour parameters were found to be $\alpha = 15$, $\lambda_{I} = 1.7$, $\lambda_{O} = 1.0$, the kernel size $s = 5$, and the range of gap parameter in the H-minima transform $\Delta \in [2, 6]$. For Dzyubachyk's method \cite{dzyubachyk10}, after carrying out the FROC analysis, parameters were fixed as $\alpha = 25, n_{i} = 50, n_{conv} = 45, n_{\varepsilon} = 0$, and the evolution time step was fixed as $\Delta t = 0.1$. For Al-Kofahi's method \cite{kofahi10}, the parameters obtained using the FROC analysis on our datasets were the minimum scale for the Laplacian of Gaussian (LoG) filter $\sigma_{\min} \in [3, 9]$, the maximum scale for the LoG filter $\sigma_{\max} \in [12, 38]$, the range of clustering resolution parameter $r \in [8, 36]$, and the range of weighting parameter for the graph cuts segmentation algorithm $\sigma_{L} \in [20, 46]$.
%----------------------------------------- FIGURE - 2 -----------------------------------------------------%
\begin{figure*}[!t]
\subfloat{\includegraphics[width= 1.75in, height=1.75in]{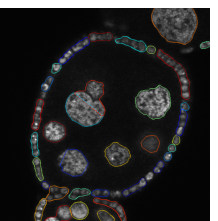}} \hfill
\subfloat{\includegraphics[width= 1.75in, height=1.75in]{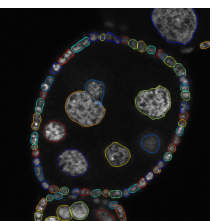}}\hfill
\subfloat{\includegraphics[width= 1.75in, height=1.75in]{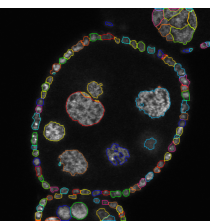}}\hfill
\subfloat{\includegraphics[width= 1.75in, height=1.75in]{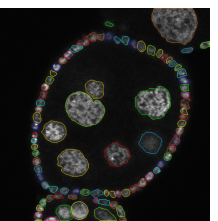}} \vspace{-2mm}\\ 
\subfloat{\includegraphics[width= 1.75in, height=1.75in]{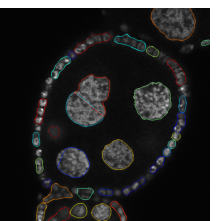}} \hfill
\subfloat{\includegraphics[width= 1.75in, height=1.75in]{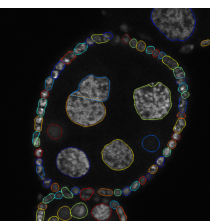}} \hfill
\subfloat{\includegraphics[width= 1.75in, height=1.75in]{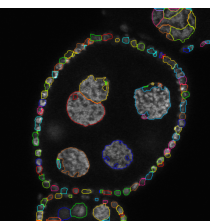}}\hfill
\subfloat{\includegraphics[width= 1.75in, height=1.75in]{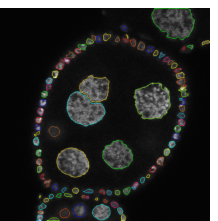}}\\
\white{0} \hfill (a) \textsc{Cheng's Method} \hfill (b) \textsc{Dzyubachyk's Method} \hfill (c) \textsc{Al-Kofahi's Method} \hfill (d) \textsc{Proposed Method} \hfill \hfill \\
\vspace{-4mm}
\caption{Comparison of segmentation results of all methods for two images of the same dataset.}
\label{fig3}
\end{figure*}
%----------------------------------------------------------------------------------------------------------------%
%---------------------------------------- TABLE - 1 -------------------------------------------------------------%
\begin{table*}[!t]
\renewcommand{\arraystretch}{2}
\caption{Mean Segmentation Accuracy (and Standard Deviation) Results}
\label{table1}
\centering
\begin{tabular}{N| M M M M M}
\hline 
 \textbf{Method} & \textbf{Mutual ~~~~ Overlap} & \textbf{Tanimoto Coefficient} & \textbf{F-score} & \textbf{Rand Index} & \textbf{Hausdorff ~~ Distance} \\ [2mm] \hline
\textbf{Proposed} & \textbf{0.9315} ~ (\textbf{0.0121}) & \textbf{0.8974} ~ (\textbf{0.0215}) & \textbf{0.9268} ~ (\textbf{0.0146}) & \textbf{0.8941}~ (\textbf{0.0084}) & \textbf{1.1926} ~ (\textbf{0.2314})\\
\textbf{Cheng} & 0.8163 ~ (0.0219) & 0.7963 ~ (0.0306) & 0.8519 ~ (0.0208) & 0.8058 ~ (0.0072) & 2.2963 ~ (0.2946)\\
\textbf{Dzyubachyk} & 0.8527 ~ (0.0244) & 0.8419 ~ (0.0186) & 0.8805 ~ (0.0107) & 0.8441 ~ (0.0092) & 1.6167 ~ (0.3646)\\
\textbf{Al-Kofahi} & 0.8631 ~ (0.0198) & 0.8387 ~ (0.0224) & 0.8498 ~ (0.0173) & 0.8495 ~ (0.0069) & 1.3968 ~ (0.1359)\\
\hline
\end{tabular}
\end{table*}
%-----------------------------------------------------------------------------------------------------------------
%--------------------------------------------- FIGURE - 3 -------------------------------------------------------- 
\begin{figure*}[!t]
\subfloat[\textsc{Original 3-D Image}]{\hspace{6mm} \includegraphics[width= 2.1in, height= 2.1in]{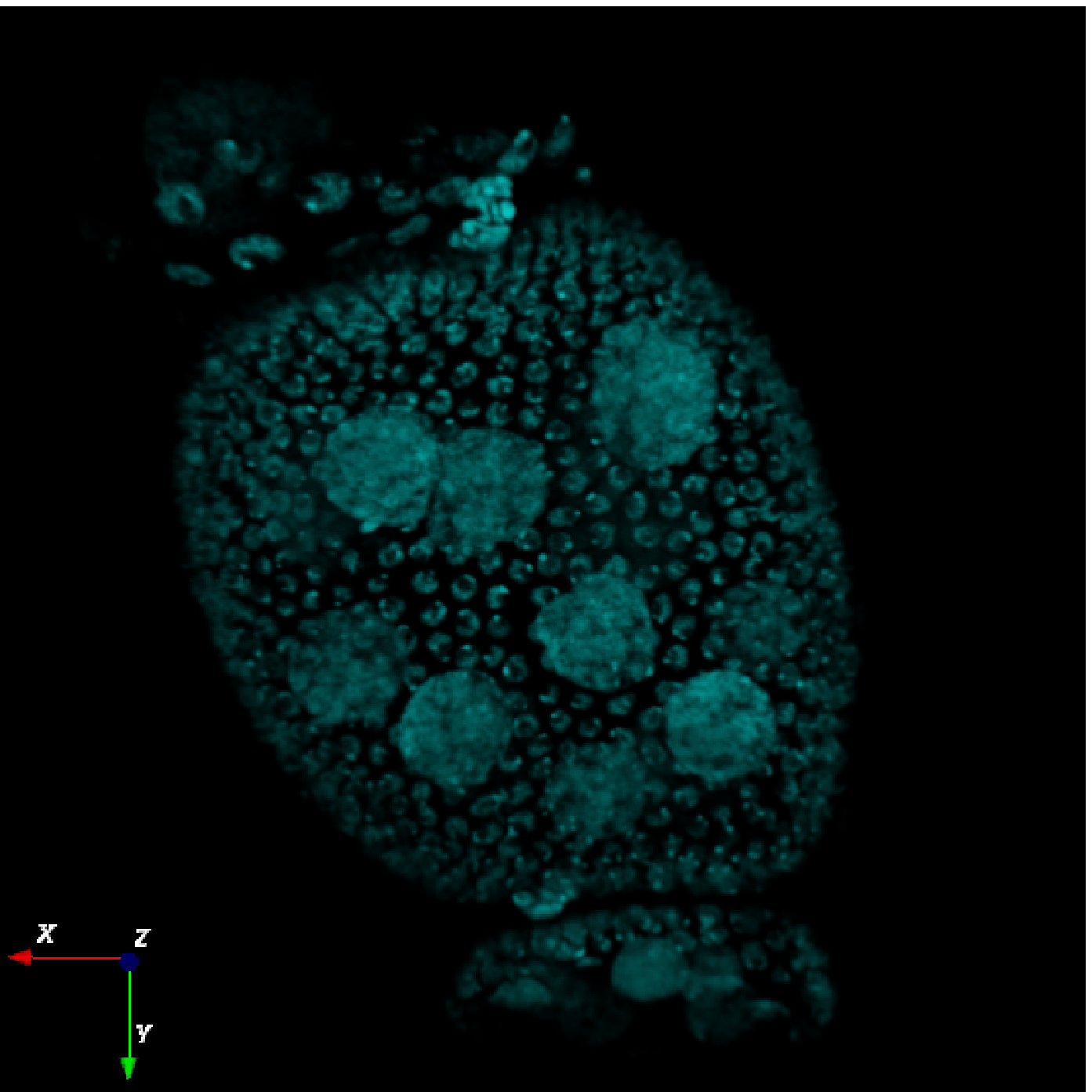}}  \hspace{2mm}
\subfloat[\textsc{Initial Segmentation}]{\includegraphics[width= 2.1in, height= 2.1in]{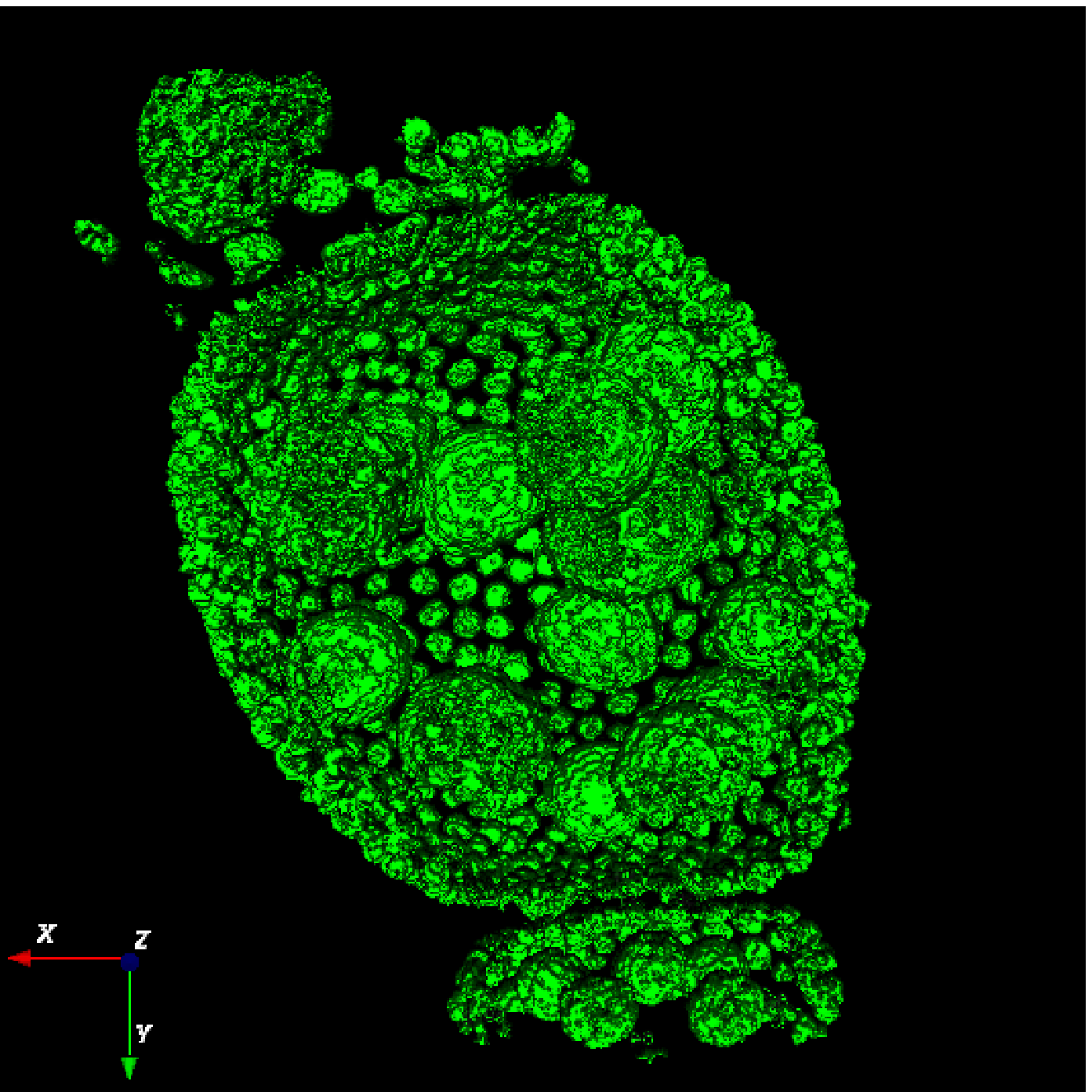}} \hspace{2mm}
\subfloat[\textsc{Final Segmentation}]{\includegraphics[width= 2.1in, height= 2.1in]{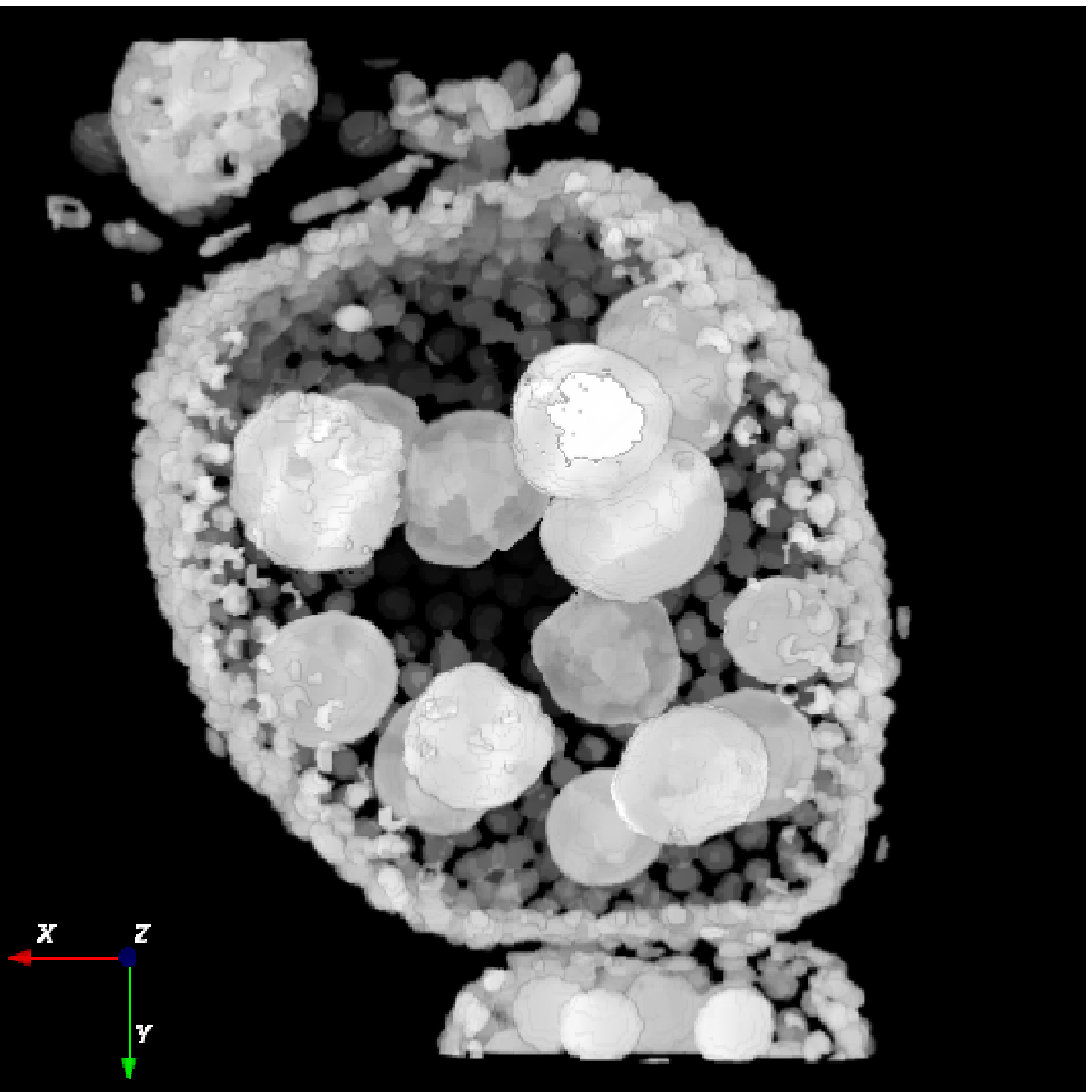}} 
\caption{Segmentation result of the proposed method on the entire 3-D data set. (a) Volume-rendered view of an example 3-D data set projected onto a coronal plane. (b) 3-D surface plot of the initial segmentation of (a). (c) Surface plot of the 3-D data set after post processing.}
\label{fig4}
\vspace{-4mm}
\end{figure*}
%----------------------------------------------------------------------------------------------------------------
%---------------------------------------- FIGURE - 4 ----------------------------------------------------------%
\begin{figure*}[!t]
\subfloat{\includegraphics[width= 7.1in, height=5in]{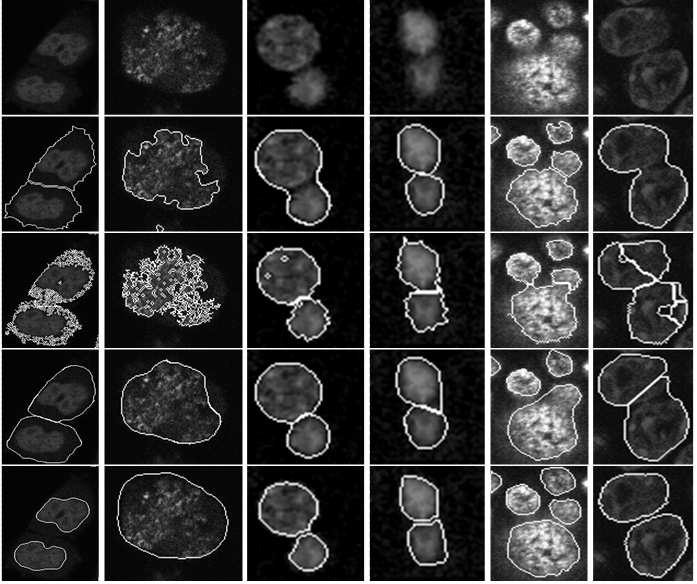}}
\vspace{-2mm}
\caption{Comparison of segmentation results of all the automated methods for various 2-D slices from different datasets. Row 1: original 2-D slice image, Row 2: Cheng's method, Row 3: Al-Kofahi's method, Row 4: Dzyubachyk's method and Row 5: proposed method.}
\label{fig5}
\vspace{-3mm}
\end{figure*}
%----------------------------------------------------------------------------------------------------------------%
\subsection{Nucleus Segmentation Evaluation}
\label{subsec:segeval}

We demonstrate the performance of the proposed algorithm and the three other segmentation algorithms, using a variety of segmentation evaluation metrics described below. Table \ref{table1} shows the segmentation accuracy for all the automated segmentation algorithms for the eight datasets used for testing.

\subsubsection{Mutual Overlap}
\label{subsubsec:dice}

This metric, also known as Dice evaluation \cite{sonka07}, is based on the mutual overlap between the automated segmentation algorithms and the carefully performed manual segmentation. Specifically, the mutual overlap is calculated using the automatically segmented region $R_{1}$ and the manually segmented region $R_{2}$ as follows:
\begin{eqnarray}
2 \frac{\left| R_{1} \cap R_{2} \right|}{\left| R_{1} \right| + \left| R_{2} \right|}
\end{eqnarray}
where $ \left| \cdot \right|$ represents the number of voxels in the region. The value is bounded between zero (no overlap) and one (exact overlap). 

\subsubsection{Tanimoto Coefficient}
\label{subsubsec:tc}

The Tanimoto coefficient \cite{duda01} measures the  similarity between the automatically segmented region $R_{1}$ and the manually segmented region $R_{2}$. For this segmentation problem, the Tanimoto coefficient is the number of voxels that regions $R_{1}$ and $R_{2}$ have in common, divided by the total number of voxels belonging to either region. It is computed as
\begin{eqnarray}
\frac{\left|{R_{1}\cap R_{2}}\right|}{\left|{R_{1}} \right| + \left|{R_{2}} \right| - \left|{R_{1}\cap R_{2}} \right|} = \frac{ \left|{R_{1}\cap R_{2}} \right|}{\left|{R_{1}\cup R_{2}} \right|}
\end{eqnarray} 
The value is bounded between zero (no overlap) and one (exact overlap). 

\subsubsection{F-score}
\label{subsubsec:fscore}

Precision ($\mathcal{P}$) and recall ($\mathcal{R}$) are two common measures for evaluating the quality of results using the automatically segmented region $R_{1}$ and the manual segmented region $R_{2}$. They are defined as
\begin{eqnarray}
\mathcal{P} = \frac{\left| R_{1}\cap R_{2} \right|}{\left| R_{1} \right|}, \quad \mathcal{R} = \frac{\left| R_{1}\cap R_{2} \right|}{\left| R_{2} \right|}
\end{eqnarray}
The F-score \cite{ram18-2} is a statistical measure of a test's accuracy that combines both precision ($\mathcal{P}$) and recall ($\mathcal{R}$) into a single measure by taking their harmonic mean:
\begin{eqnarray}
2 \frac{\mathcal{P} \mathcal{R}}{\mathcal{P} + \mathcal{R}}
\end{eqnarray}
The F-score is bounded between zero (bad segmentation) and one (perfect segmentation).

\subsubsection{Rand Index}
\label{subsubsec:randerr}

The Rand index is a well-known measure of similarity between two data clusterings. It was proposed as a measure of clustering performance by W. M. Rand \cite{rand71}. Unnikrishnan \emph{et al.} \cite{unnikrishnan07} proposed the normalized probabilistic Rand (NPR) index. The NPR normalizes the Rand index using a baseline that assumes that the segmentations are generated from a hypergeometric distribution. The significance of this measure is called into question by Meila \cite{meila07}, who notes that this measure makes strong assumptions about the distribution, and that some pairs of clusterings may result in negative index values. Also, computing the expected index values can be very computationally expensive \cite{unnikrishnan07}. For these reasons we avoid the NPR index. The Rand index is defined as a measure of agreement. Given the manually segmented region $R_{1}$ and automatically segmented region $R_{2}$ of a 3-D image $I$ with $n$ voxels, let $a$ denote the number of (unordered) pairs of voxels in $I$ that are in the same object in $R_{1}$ and also in the same object in $R_{2}$, and let $b$ denote the number of pairs of voxels in $I$ that are in different objects in $R_{1}$ and in different objects in $R_{2}$. The Rand index is then defined as the frequency with which two segmentations agree on whether a pair of voxels belongs to the same or different objects:
\begin{eqnarray}
\frac{{(a + b)}}{\binom{n}{2}} 
\end{eqnarray}
The value is bounded between zero (bad segmentation) and one (perfect segmentation).

\subsubsection{Hausdorff Distance}
\label{subsubsec:hdist}

This measures the extent to which each voxel on the boundary of a manually segmented region $R_{1}$ lies near some voxel on the boundary of an automatically segmented region $R_{2}$ and vice-versa. Thus, it can be used to determine the degree of resemblance of two segmented regions. It is defined as the greatest of all the distances from points in one set to the closest points in the other set. If we consider the manually segmented region $R_{1} = \{ r_{1}, r_{2}, \dots , r_{m} \}$ and automatically segmented region $R_{2} = \{ r'_{1}, r'_{2}, \dots , r'_{n} \}$ where $r_{1}, r_{2}, \dots , r_{m}$ and $r'_{1}, r'_{2}, \dots , r'_{n}$ are the boundary voxels of region $R_{1}$ and $R_{2}$, respectively, then the Hausdorff distance \cite{huttenlocher93} is defined as
\begin{eqnarray}
 \max \left \{ h(R_{1}, R_{2}), h(R_{2}, R_{1}) \right \}
\end{eqnarray}
where
\begin{eqnarray*}
h(R_{1}, R_{2}) = \max_{r \in R_{1}} \, \min_{r' \in R_{2}} \parallel r - r' \parallel
\end{eqnarray*}
is called the \emph{directed Hausdorff distance} from set $R_{1}$ to $R_{2}$ where $\parallel \cdot \parallel$ is the Euclidean distance between the boundary voxels of $R_{1}$ and $R_{2}$, and
\begin{eqnarray*}
h(R_{2}, R_{1}) = \max_{r' \in R_{2}} \, \min_{r \in R_{1}} \parallel r' - r \parallel
\end{eqnarray*}
is called the \emph{reverse Hausdorff distance}. A Hausdorff distance of zero signifies that there is a perfect overlap between the boundaries of the two segmented regions. Thus, lower Hausdorff distance means better segmentation.   

Fig. \ref{fig3} shows example segmentation results obtained for two representative 2-D slices from one of our 3-D data set for the four methods under consideration. Fig. \ref{fig4} shows the segmentation result of our proposed method for the whole 3-D data set viewed as 3-D surface plot. To further show the robustness of the proposed cell nuclei segmentation algorithm, we ran the four automated algorithms considered in this paper on the publicly available histopathological 3-D cell image data reported as a benchmark for cell tracking \cite{maska14}. Fig. \ref{fig5} shows example segmentation results obtained for some representative 2-D slices from the 3-D datasets available online.\footnotemark \footnotetext{\url{http://www.codesolorzano.com/celltrackingchallenge/Cell_Tracking_Challenge/Datasets.html}}
 %------------------------------------------- TABLE - 2 ---------------------------------------------------------
\begin{table}[!t]
\caption{Mean Segmentation Accuracy (and Standard Deviation) Results With and Without Post Processing}
\label{table2}
\begin{center}
\renewcommand{\arraystretch}{2}
\begin{tabular}{>{\centering} m{1.4cm}| >{\centering} m{1.2cm}  >{\centering}m{1.5cm} >{\centering} m{1.2cm} >{\centering}m{1.3cm}}
\hline 
 \textbf{Proposed Method} & \textbf{Mutual Overlap} & \textbf{Tanimoto Coefficient} & \textbf{F-score} & \textbf{Rand Index} \tabularnewline
 \hline
\textbf{W/ Post Processing} & \textbf{0.9315} (\textbf{0.0121}) & \textbf{0.8974} (\textbf{0.0215}) & \textbf{0.9268} (\textbf{0.0146}) & \textbf{0.8941}  (\textbf{0.0084}) \tabularnewline
\textbf{W/O Post Processing} & 0.9096 (0.0192) & 0.8692 (0.0102) & 0.9001 (0.0153) & 0.8622  (0.0163) \tabularnewline
\hline
\end{tabular}
\end{center}
\vspace{-5mm}
\end{table}
%-----------------------------------------------------------------------------------------------------------------

\subsection{Effect of Post Processing}
\label{subsec:effectpp}

In order to show the effect of post processing step in the algorithm, we ran the algorithm on all the eight 3-D datasets with and without the post processing step, and computed the segmentation accuracy metrics in both cases. Table II shows the mean and standard deviation of the segmentation accuracy metrics with and without post processing. From Table II, we observe that the mutual overlap, Tanimoto coefficient, F-score and the Rand index of the proposed method with post processing is greater than that of the proposed method without post processing. This shows that post processing is important in terms of increasing the accuracy of the proposed method.

\subsection{Speed}
\label{subsection:speed}

We also compare the proposed automated algorithm to the other three methods in terms of speed. We ran all the algorithms on the eight datasets. Our proposed method and Cheng's method \cite{cheng09} were coded in MATLAB. For Dzyubachyk's method \cite{dzyubachyk10} and Al-Kofahi's method \cite{kofahi10}, we obtained the research source code available online, implemented in MATLAB. The whole process was carried out on a 2.67 GHz, 6 GB RAM, Intel Core i5 processor, Windows PC. Table \ref{table3} shows mean computation time (and standard deviation) for the four methods.

\section{Discussion and Conclusion}
\label{sec:diss}

We have proposed a new automated technique for detection and segmentation of 3-D nuclei from FISH images. Using FISH data from Drosophila cells, we compared the proposed algorithm with three other techniques with respect to various performance metrics. Evaluation experiments were performed on thirty eight fluorescence microscopy datasets consisting of 8519 follicle cells and 584 nurse cells, so as to be representative of a variety of biological data.
  
Nucleus segmentation evaluation was carried out with respect to five different metrics. As shown in Table \ref{table1}, the mutual overlap accuracy of the proposed algorithm is 11.52 percentage points greater than that of Cheng's method, 7.88 percentage points greater than Dzyubachyk's method, and 6.84 percentage points greater than Al-Kofahi's method. The Tanimoto coefficient of the proposed algorithm is 10.11 percentage points greater than Cheng's method, 5.55 percentage points greater than Dzyubachyk's method, and 5.87 percentage points greater than Al-Kofahi's method. The F-score of the proposed method is 7.79 percentage points greater than Cheng's method, 4.93 percentage points greater than Dzyubachyk's method, and 8 percentage points greater than Al-Kofahi's method. We observe from Table \ref{table1} that the Rand index of 0.8941 for the proposed method is the largest in comparison to 0.8058--0.8495 for the other three methods. Also, Table \ref{table1} shows that the Hausdorff distance for the proposed method 1.1926 is the smallest compared to 1.3968--2.2963 for the other three methods
 %------------------------------------- TABLE - 3 -----------------------------------------------------------
\begin{table}[!t]
\caption{Mean Computation Time (and Standard Deviation) (in MM:SS)}
\label{table3}
\begin{center}
\renewcommand{\arraystretch}{2}
\begin{tabular}{>{\centering} m{1.5cm}| >{\centering} m{1.3cm}  >{\centering}m{1.2cm} >{\centering} m{1.3cm} >{\centering}m{1.3cm}}
\hline 
 \textbf{Method} & \textbf{Proposed} & \textbf{Cheng} & \textbf{Dzyubachyk} & \textbf{Al-Kofahi} \tabularnewline
 \hline
\textbf{Computation Time} & \textbf{06:26} (\textbf{01:07}) & 21:06 (01:01) & 47:29 (02:10) & 03:23  (00:49) \tabularnewline
\hline
\end{tabular}
\end{center}
\vspace{-5mm}
\end{table}
%-----------------------------------------------------------------------------------------------------------------

The proposed algorithm was also evaluated with respect to computation time. From Table \ref{table3} we see that the proposed algorithm runs 3 minutes and 3 seconds slower than Al-Kofahi's method, 14 minutes and 46 seconds faster than Cheng's method and 41 minutes and 3 seconds faster than Dzyubachyk's method. This automated technique lays the foundation for further analysis of 3-D spatial organization of sub-nuclear structures and genes within the cells. Furthermore, the proposed method was developed for Drosophila cell nuclei in the context of whole mount tissue FISH, and, therefore, this method may have general application to thick tissue samples where many optical slices must be considered in order to reconstruct a complete 3-D image. 

\section*{Acknowledgment}

We thank Prof. Giovanni Bosco (Dept. of Genetics, Dartmouth College) for useful conversations and insights on the structure of ovarian germline cells and FISH imaging and for providing the fluorescence microscopy image datasets used in our study.   

\ifCLASSOPTIONcaptionsoff
  \newpage
\fi

\IEEEtriggeratref{26}


%\bibliographystyle{IEEEtran}
\bibliographystyle{IEEEtran}
\bibliography{IEEEabrv,myref}
\end{document}